\documentclass[12pt,preprint]{aastex}

\newcommand{\afe}{[\alpha/\mathrm{Fe}]}
\newcommand{\Ms}{M_{\odot}}

\slugcomment{To appear in ApJ Letters}

\shorttitle{HB Morphology and Mass Loss}
\shortauthors{A. Dotter}

\begin{document}

\title{Horizontal Branch Morphology and Mass Loss in Globular Clusters}

\author{Aaron Dotter}
\affil{Department of Physics and Astronomy, University of Victoria,\\
    P. O. Box 3055, Victoria, British Columbia, V8W 3P6, Canada}
\email{dotter@uvic.ca}

\begin{abstract}
The connection between mass loss on the red giant branch (RGB) and horizontal
branch (HB) morphology in globular clusters (GCs) has long been acknowledged
but the mechanisms governing mass loss remains poorly understood from a theoretical 
perspective.  The present study uses synthetic HB models to demonstrate for the first 
time that $\alpha$-enhancement and a simple relation between mass loss and metallicity
can explain the entire range of HB morphology (characterized by the HB type index) 
observed in old, coeval GCs.  The mass loss-metallicity relation accounts naturally 
for the fact that the most metal poor GCs ([Fe/H] $< -2$) have redder HBs than is 
typical of GCs with $-2 <$ [Fe/H] $< -1.5$ without invoking younger ages. These results
may prove useful in studying the contribution of HB stars to integrated light via
stellar population synthesis.
\end{abstract}


\section{Introduction}
\citet{z93} recommended that Galactic halo GCs be divided into two groups: 
one for old, roughly coeval GCs that follow the mean trend in [Fe/H] vs. HB 
type in the inner halo (the Old Halo; OH) and one for GCs that deviate from 
this trend by a significant amount (the Younger Halo; YH).  Due to the lack 
of very metal poor GCs in the inner halo, Zinn used more distant GCs to
extend the line below [Fe/H] $< -2$, see his Figure 1.

Zinn's relation was empirically motivated and differs from later studies---such 
as those by \citet[hereafter LDZ]{ldz2} and \citet{re01} that are based 
on synthetic HB isochrones---in that it turns back to the red at the lowest 
metallicities ([Fe/H] $<$ -2). Trends derived from synthetic HB calculations 
with constant mass loss (LDZ) or mass loss that increases with [Fe/H] and age 
\citep{re01} indicate that HB type (as defined by LDZ 1994) rises monotonically 
to +1 as [Fe/H] decreases and remains there for all lower values of [Fe/H].  
\citet{l91} showed that allowing mass loss to vary with [Fe/H], in a manner 
consistent with the observed trend in GCs, resulted in better agreement between 
his models and the pulsational properties of RR Lyrae stars in $\omega$ Cen than 
was achieved by assuming constant mass loss. Aside from Lee's application to 
$\omega$ Cen, the concept of searching for a mass loss relation that matches the 
observed trend in old, coeval GCs has not been explored.

The purpose of this Letter is to argue that it is possible to account for
the fact that the most metal poor GCs have HB types less than +1 without
requiring them to be significantly younger than more metal rich GCs if
mass loss on the RGB varies as a function of [Fe/H] in a manner qualitatively
similar to that proposed by \citet{l91}.

\section{Synthetic HB Model Construction}
\citet{ro73} introduced an effective means of modeling the distribution 
of HB stars in GCs by assuming the stars that arrive on the HB are drawn at 
random from a Gaussian mass distribution that is truncated on both ends by 
physical limitations: namely the initial mass and the He core mass of the 
precursor red giant stars.  The method further assumes that stars arrive on 
the HB at a constant rate and spend roughly the same amount of time there 
regardless of mass.  Under these assumptions, one constructs a synthetic 
HB by drawing mass and age at random and then interpolating in a grid of 
He-burning evolutionary tracks to obtain the relevant quantities.

Many investigators, such as \citet{ro73}, LDZ (1990, 1994), \citet{ca94},
\citet{re01}, and \citet{ca04}, have used the synthetic HB method to show how
various model parameters influence HB morphology and the pulsational properties 
of RR Lyrae stars in GCs.
LDZ (1994) generated isochrones in the [Fe/H] vs. HB type diagram for different 
ages and He abundances while assuming a fixed amount of mass loss and dispersion. 
\citet{re01} incorporated the Reimers mass loss formula into their synthetic 
HB analysis (such that the average mass loss increased with [Fe/H] and age) and 
found that, while the sensitivity of HB morphology increased compared to the 
findings of LDZ (1994), the general trend was unchanged. These isochrones were 
used to argue that GCs with anomalously red HBs (compared to the oldest synthetic 
HB isochrone) are younger than the OH GCs: one of strongest arguments in favor of 
age as the second parameter governing HB morphology in GCs.

This study employs the He core masses, RGB tip masses, and He-burning 
tracks from \citet{d07} to create synthetic HB models (each consisting of 500
stars) spanning the range of [Fe/H] and $\afe$ observed in the Galactic halo. 
Simulations of 1000-2000 synthetic HBs are generated by allowing [Fe/H] to 
vary between $-2.5$ and $-0.5$ while $\afe$, age, mass loss ($\Delta$M), and 
dispersion ($\sigma_M$) are held constant or varied in a specified manner.  

\begin{figure}
\plotone{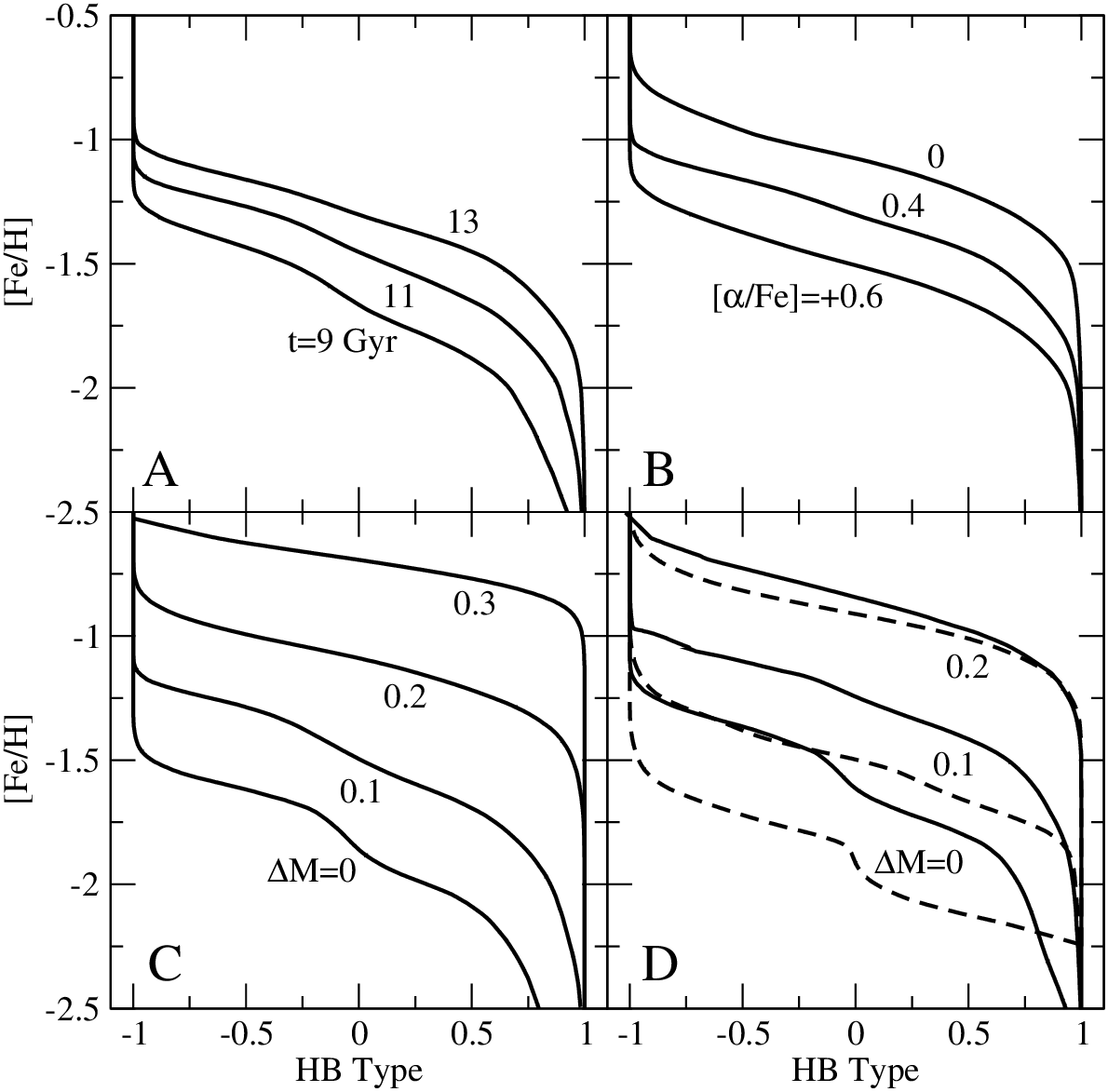}
\caption{The effect of varying: (A) age at constant $\Delta$M=0.15, $\sigma_M$=0.02,
 and $\afe$=+0.4; (B) $\afe$ at constant $\Delta$M=0.15, $\sigma_M$=0.02, and 13 
Gyr. (C) $\Delta$M at constant $\sigma_M$=0.02, age of 13 Gyr, and $\afe$=+0.4; and 
(D) $\Delta$M at constant $\sigma_M$=0.02 using the He-burning tracks of 
\citet[dashed line; Y=0.23, $\afe$=0, 13 Gyr]{ca04} and \citet[solid line; 
Y=0.245-0.264, $\afe$=0, 13 Gyr]{d07}.\label{param}}
\end{figure}

Figure \ref{param} indicates how some of the governing parameters influence 
the [Fe/H] vs. HB type trend.  Panel A compares trend lines derived from a 
simulation with constant mass loss ($\Delta$M=0.15 $\Ms$), dispersion
($\sigma_M$=0.02 $\Ms$), and $\afe$=+0.4 for ages of 9, 11, and 13 Gyr; panel B
shows the same mass loss and dispersion assumed in A but at a fixed age (13 Gyr) 
with $\afe$=0, +0.2, +0.4, and +0.6; panel C illustrates the effect of varying 
the mass loss ($\Delta$M=0, 0.1, 0.2, 0.3 $\Ms$) when the dispersion (0.02 $\Ms$), 
age (13 Gyr), and $\afe$ (+0.4) are held constant; finally, panel D compares 
synthetic HBs derived from the He-burning evolutionary tracks of \citet[solid 
lines]{d07} and \citet[dashed lines]{ca04} at 13 Gyr with $\Delta$M=0, 0.1, and 
0.2 $\Ms$. The \citet{ca04} models were calculated at Y=0.23 with a scaled-solar 
mixture for all [Fe/H] while the \citet[see their Table 4]{d07} models range from 
Y=0.245 (at [Fe/H]=$-2.5$, $\afe$=0) to Y=0.264 (at [Fe/H]=$-0.5$, $\afe$=0) so it 
is not surprising that the the \citet{d07} models yield consistently bluer HB types 
in most regions of the diagram. 
Figure \ref{param} shows that there is considerable degeneracy in the [Fe/H] vs. HB 
type diagram among age, $\afe$, and $\Delta$M.  Though not shown,
stochastic effects are also significant because even though each synthetic HB 
consists of 500 stars (a number observed only in rich GCs) there is still a range 
of possible outcomes ($\Delta$HB type$\sim$0.1) for the same input parameters. 
Only a few tens of HB stars have been observed in some GCs and this, combined
with stochastic effects, can make the parameters governing HB morphology 
difficult to ascertain from model comparisons.

\begin{figure}
\plotone{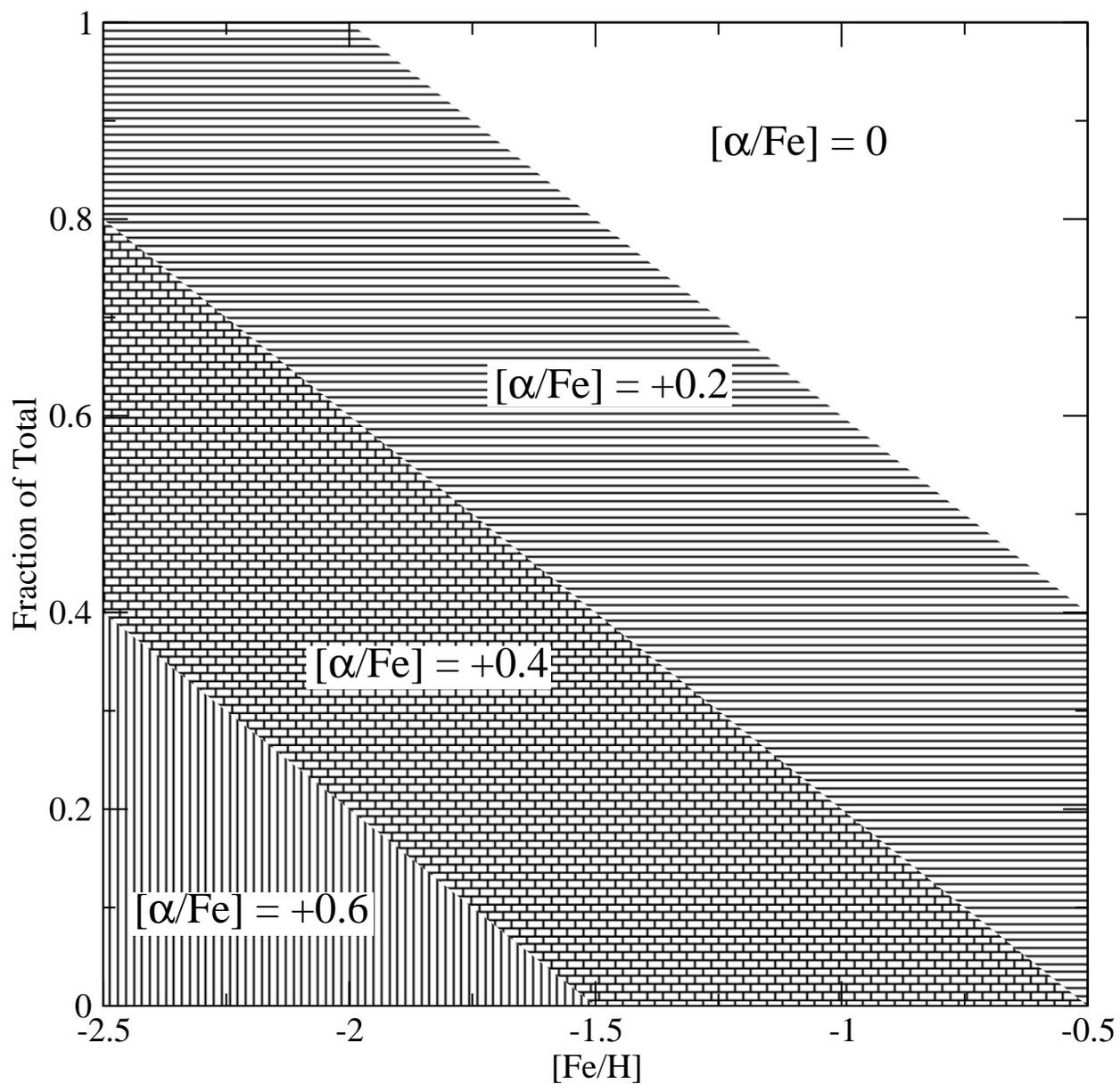}
\caption{The approximate fraction of all synthetic HBs with a given value of 
$\afe$ as a function of [Fe/H] in the synthetic HB simulations.  Since the 
number of realizations in each simulation was finite, the actual fractions 
only approach the values shown here.
\label{alpha}}
\end{figure}

\begin{figure}
\plotone{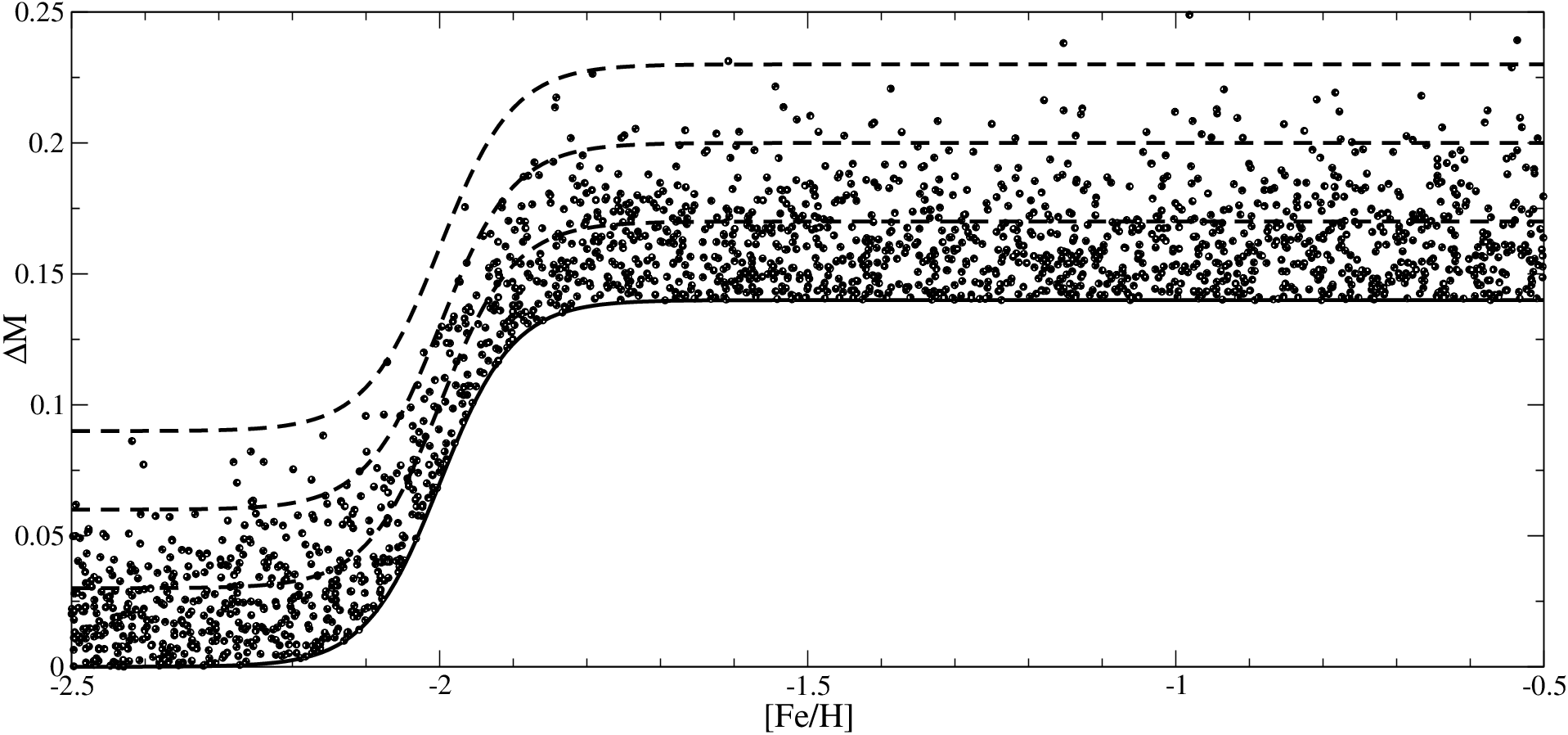}
\caption{Distribution of $\Delta$M (in $\Ms$) as a function of [Fe/H]. The points are 
from one simulation of 2000 synthetic HB models.  The simulation used $a=-2$, $b=0.05$, 
and $c=0.14$ in equation \ref{dM}. The solid line is the minimum trend 
(equation \ref{dM}) and the dashed lines are 1, 2, and 3$\sigma$ above the minimum.
\label{delsig}}
\end{figure}

Based on the relationships that were found by varying individual parameters 
in the synthetic HB models, additional simulations were generated
in order to attempt to understand the observed distribution of old,
roughly coeval GCs in the Galactic halo. In these simulations, [Fe/H] 
was allowed to vary over $-2.5 <$ [Fe/H] $< -0.5$ while age was held fixed at
13 Gyr. The level of $\afe$ was drawn at random from a flat distribution in 
such a way that the average value of $\afe$ decreased as [Fe/H] increased.  
Figure \ref{alpha} shows how the  fraction of synthetic HB models with a given 
value of $\afe$ is assumed to vary with [Fe/H].  
This method of distributing $\afe$ is based on observations of the 
$\alpha$-elements as a function of [Fe/H] in halo GCs and field stars 
\citep[see their Figure 3]{pvi05}.
Mass loss ($\Delta$M; measured in $\Ms$) was assumed to vary with [Fe/H] as
shown in Figure \ref{delsig}. The minimum value was set as a function of [Fe/H] 
as in equation \ref{dM}.

\begin{equation}
\Delta M_{min}(x)=c\frac{exp(\frac{x-a}{b})}{1+exp(\frac{x-a}{b})}\label{dM}
\end{equation}

Where $x$=[Fe/H] and $a$, $b$, and $c$ are free parameters.
The variation of $\Delta$M with [Fe/H] could also have been described by a 
broken linear relation; the use of equation \ref{dM} was chosen because it provides
a smooth variation with [Fe/H].  With the minimum value of $\Delta$M set by equation 
\ref{dM}, the actual value for a single synthetic HB realization $i$ was set by 
equation \ref{dMR}.

\begin{equation}
\Delta M_i = \Delta M_{min}(x_i) + |R_G(0,0.03)|\label{dMR}
\end{equation}

Where $R_G(y,z)$ is a random number drawn from a Gaussian mass distribution having a 
mean value $y$ and dispersion $z$.  Thus the value of $\Delta$M was more likely to 
lie near the minimum value but, given a large enough simulation, values at the 
$\sim$3 $\sigma$ level can be found. Figure \ref{delsig} shows the distribution of 
$\Delta$M as a function of [Fe/H] using equations \ref{dM} and \ref{dMR} for a 
simulation of 2000 synthetic HB models. The minimum value as defined by equation 
\ref{dM} is shown as the solid line while the dashed lines represent the minimum 
value plus 1, 2, and 3$\sigma$.

The value of the dispersion in the mass distribution ($\sigma_M$; also measured in 
$\Ms$) was set in a similar fashion to that for $\Delta$M (equation \ref{dMR}) except 
that the minimum value was held constant at 0.02 and the dispersion in $R_G$ was 0.01.

\section{Results}

\begin{figure}
\plotone{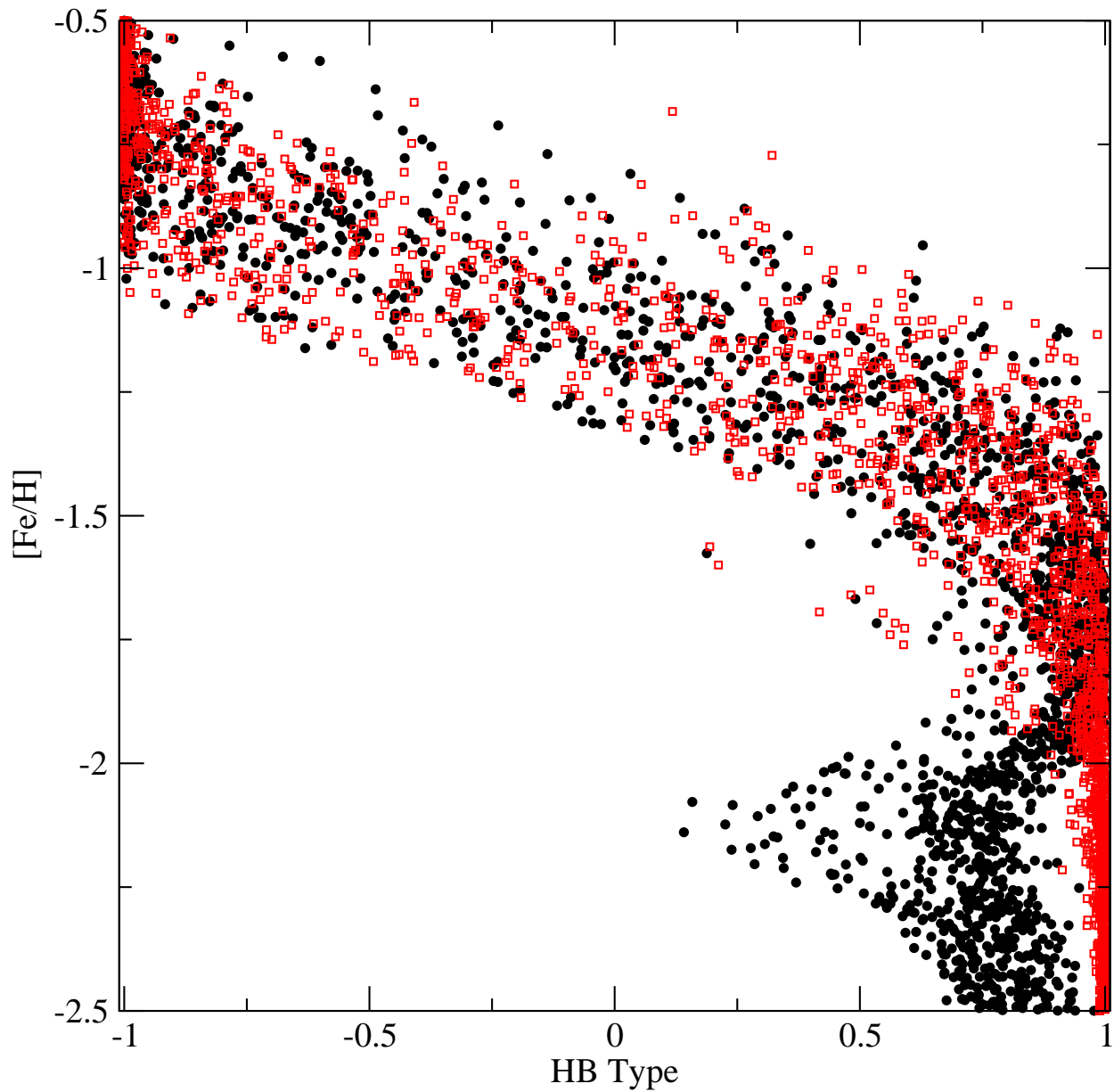}
\caption{Comparison of synthetic HB simulations for which $\Delta$M$_{min}$ was 
either fixed (open squares) or varied according to equation \ref{dM}
(solid circles).  In both cases, $\Delta$M was set by equation \ref{dMR}.
HB type varies monotonically with [Fe/H] in the fixed $\Delta$M$_{min}$ case but 
turns back to the red in the variable $\Delta$M$_{min}$ case below [Fe/H] $\la -2$.
\label{deltam}}
\end{figure}

Figure \ref{deltam} shows the [Fe/H] vs. HB type diagrams that are obtained 
when the variable $\Delta$M$_{min}$ given by equation \ref{dM} with $a=-2, 
b=0.05$, and $c$=0.14 (solid circles) and a constant $\Delta$M$_{min}$=0.14 
(open squares) are assumed. Both simulations consist of 2000 synthetic HB models at 
13 Gyr and both used equation \ref{dMR} to determine $\Delta$M and $\sigma_M$ as 
outlined in the preceding section. That synthetic HB models become redder as $\afe$ 
increases at a given [Fe/H] may be a necessary part of the explanation for why at 
least some of the most metal poor GCs have redder HBs than can be found at [Fe/H] 
$\ga -2$ but $\afe$ alone is not sufficient to explain this behavior.  Assuming a 
constant mass loss and the $\afe$ distribution shown in Figure \ref{alpha} produces 
the same sort of [Fe/H] vs. HB type behavior demonstrated by LDZ (1994) and \citet{re01}.  
Allowing $\Delta$M to vary with [Fe/H] (see equations \ref{dM} and \ref{dMR}) as 
described above produces a trend that matches the original OH trend of \citet{z93}.

\begin{figure}
\plotone{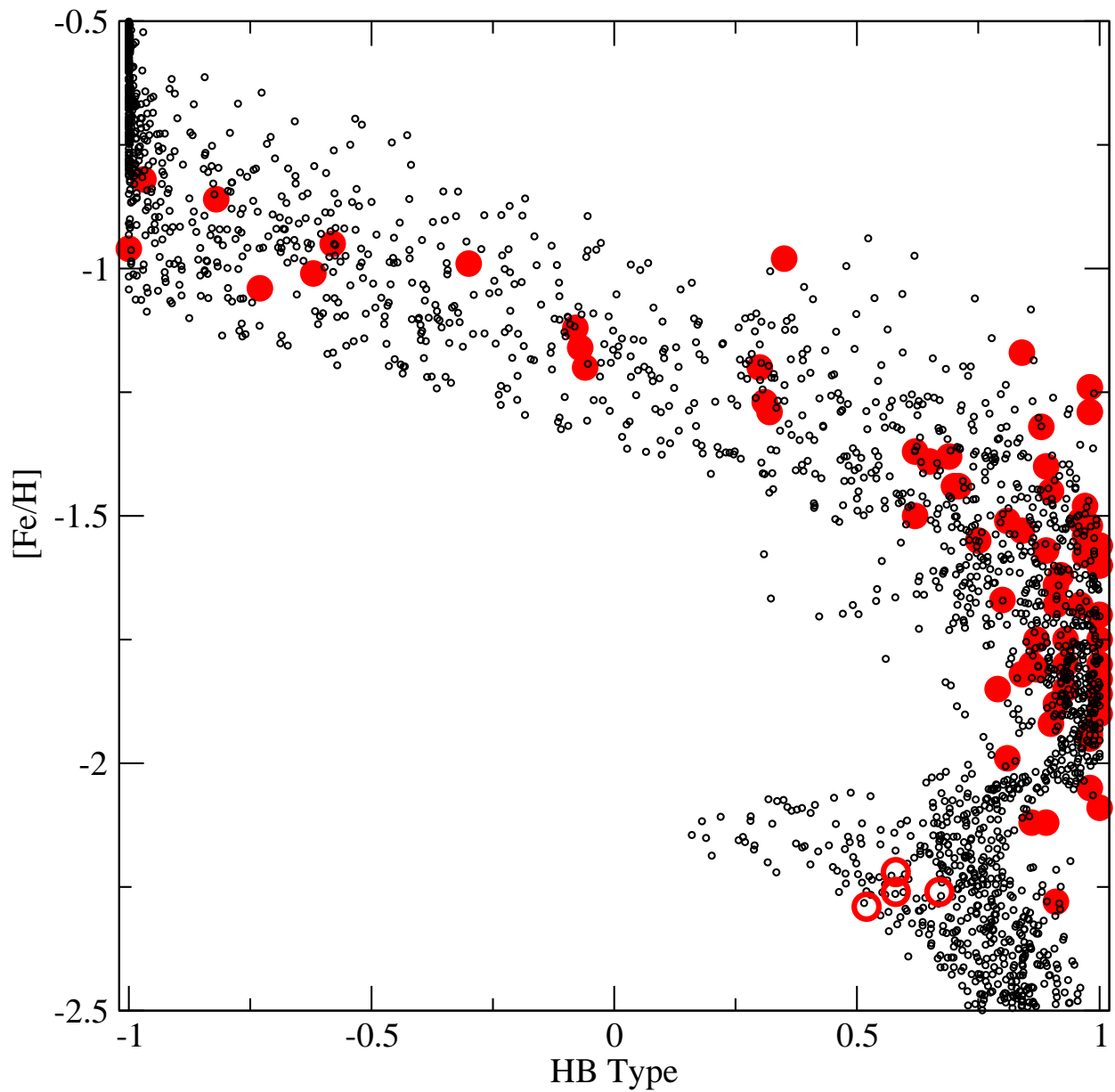}
\caption{The synthetic HB simulation with variable mass loss (small filled cirlces) 
compared to the OH GC sample (large filled circles) and four YH GC (large open circles)
of \citet{mvdb} with omissions as described in the text. The $\Delta$M-[Fe/H] relation 
given by equation \ref{dM} is capable of simultaneously reproducing the OH GCs and the 
four metal poor YH GCs without assuming any variation in age.\label{gcshb}}
\end{figure}

The GC catalog compiled by \citet{mvdb} places GCs into sub-populations 
along the lines of \citet{z93} but with additional categories that reflect 
discoveries made in the intervening years. \citet{mvdb} used the HB isochrones 
of \citet{re01} to assign a sub-population  to each GC. Many of the lowest 
metallicity GCs are considered part of the YH despite evidence that their
ages are similar to other metal poor OH GCs, such as M\,92 and NGC\,2419, with
bluer HBs \citep{vdb00,sw02}.  Given the lack of a substantial
age difference between the lowest metallicity GCs with different HB types, the
metal poor GCs NGC\,5053, 5466, 6426, and 7078 (M\,15) have been included in 
the comparison presented here.
Figure \ref{gcshb} compares a synthetic HB simulation with $\alpha$-enhancement
and variable mass loss with the OH GCs (filled circles) and four metal poor YH 
GCs (open circles) from \citet{mvdb}. In order to avoid complications, as much as 
currently possible, that arise from the HB morphologies of GCs with multiple 
stellar populations, NGC\,1851 \citep{mi08} and NGC\,2808 \citep{pi07} are not 
shown in Figure \ref{gcshb}.

The variable mass loss suggested in the previous section reproduces the distribution 
of old GCs in the halo.  Indeed, if the He-burning tracks of \citet{ca04} are used 
instead of the \citet{d07} tracks, approximating $\alpha$-enhancement
by [M/H] $\simeq$ [Fe/H] + log(0.638\,x\,10$^{\afe}$ + 0.362) \citep{scs93}, variable
mass loss with the same values of $a$, $b$, and $c$ listed at the beginning of this 
section produces a nearly identical result to that shown in Figure \ref{gcshb}.
Based on these results pertaining to mass loss and HB morphology as well as the 
evidence that the four old, metal poor GCs with relatively red HBs are not especially 
young \citep{vdb00,sw02}, it is recommended that those GCs be considered part of the
OH as originally intended by \citet{z93}.

\section{Discussion and Conclusion}

\citet{ca05} provides a thorough discussion of mass loss and HB morphology in GCs.  
Catelan considers a handful of empirically motivated formulae that describe the mass 
loss rate as a function of stellar parameters such as luminosity, radius, and surface 
gravity. Each predicts that mass loss increases smoothly with
[Fe/H] \citep[Figure 4]{ca05}. Some of the formulae are consistent with the observed 
difference in HB morphology between second parameter GC pairs, but not all, and no 
single formula is successful for each of the GC pairs considered \citep[Figures 
10-13]{ca05}. There is also a discrepancy between the formulae and observational 
efforts to measure mass lost by red giants in GCs by \citet{or02} who 
found that mass loss is not strongly correlated with metallicity and only occurs 
near the tip of the RGB.  On the other hand, \citet{or07} found that mass loss is 
evident at the level of HB and increases with luminosity up to the RGB tip
in 47\,Tuc.  The proposed mass loss-[Fe/H] relation should be straightforward 
to test observationally with mass loss estimates for a large enough GC sample.

These examples illustrate that mass loss remains a complex and poorly understood
process.  It is by no means guaranteed that the information provided by standard
stellar evolution models is sufficient to calculate mass loss as a function of
age and metallicity.  The observed range of HB morphologies in OH GCs (which are
assumed to be roughly coeval) of similar metallicity suggests that more than
one factor must be involved.  It seems quite plausible that mass loss may be 
influenced at the stellar level (by, metallicity and rotation, for example) and 
at the cluster level (by cluster mass or concentration, for example).  Uncertainties 
in composition, coming from the lack of accurate abundance determinations in some 
lesser known GCs, or the possible presence of multiple populations with distinct 
compositions in a single GC, complicate the analysis and interpretation.

The situation is further complicated by the discovery of bi- and multi-modal HBs
in some GCs, see \citet{fe98} for examples and a review of the subject.  It has
been suggested by, e.g. \citet{da08} and references therein, that these (and perhaps
all) GCs are comprised of chemically distinct populations that create the complex 
distribution of HB stars.  Complications arise even in GCs with uni-modal HBs.
The analysis of M\,3 by \citet{va08} indicates that while the synthetic HB method 
developed by \citet{ro73}, particularly the assumed Gaussian mass distribution, is 
`a reasonable first approximation, [it] fails to account for the detailed shape of 
M\,3's HB mass distribution.'  \citet{va08} also warn that there is a mismatch 
between observed and theoretical lifetime-luminosity relations on the HB.

In the absence of a firm theoretical foundation for mass loss in stars, it is 
still possible to further our understanding of this subject and to apply that 
knowledge.  One application of the results presented here is to the modeling of 
integrated light via stellar population synthesis.  Investigators such as 
\citet{le00} and \citet{le02} have explored the dependence of integrated light 
properties on synthetic HB morphology and compared them with observations of 
extragalactic GCs.  The results presented here may provide useful input for future 
studies along these lines.

To summarize, synthetic HB models that include the effects of $\alpha$-enhancement
and metallicity-dependent mass loss (via equations \ref{dM} and \ref{dMR}) can 
account for the entire distribution of HB morphologies observed in old, coeval GCs 
in the Galactic halo. Essentially the same result is obtained from two distinct sets 
of He-burning tracks, suggesting this is not a spurious or model-dependent result.  
As a corollary, it is recommended that NGC\,5053, 5466, 6426, and 7078, recently 
associated with the YH, be considered members of the OH since their ages are 
consistent with other OH GCs and their HB morphologies can be explained without
invoking substantially younger ages.

\acknowledgments

Thanks to A. Sarajedini for the motivation to pursue this work and to D. 
VandenBerg for helpful discussions and comments on the manuscript.
Thanks also to the anonymous referee.
This work was supported by a CITA National Fellowship to AD.
Support from an NSERC grant to D. VandenBerg is also acknowledged.

\clearpage

\end{document}